\documentclass[compsoc,conference,a4paper,10pt,times]{IEEEtran}
\IEEEoverridecommandlockouts
\usepackage{cite}
\usepackage{amsmath,amssymb,amsfonts}
\usepackage{algorithmic}
\usepackage{graphicx}
\graphicspath{./Downloads/}
\usepackage{textcomp}
\usepackage{bmpsize}
\usepackage{xcolor}
\usepackage{lipsum}
\usepackage[colorlinks=true,urlcolor=black]{hyperref}
\def\BibTeX{{\rm B\kern-.05em{\sc i\kern-.025em b}\kern-.08em
    T\kern-.1667em\lower.7ex\hbox{E}\kern-.125emX}}

\usepackage{booktabs} 
\usepackage{versions}
\excludeversion{RedundantContent}

\usepackage[english]{babel}
\usepackage[latin1]{inputenc}
\usepackage{amsmath,amsthm, amssymb, latexsym}
\usepackage{xspace}

\usepackage{import}
\usepackage{graphicx}
\usepackage{multirow}
\usepackage{booktabs}
\usepackage{textgreek}
\usepackage[neverdecrease]{paralist}

\usepackage{balance}
\usepackage{float}
\usepackage{caption}
\usepackage{subcaption}

\usepackage{rotating}

\usepackage{balance}
\usepackage{tcolorbox}

\usepackage{amssymb}
\usepackage{pifont}
\usepackage[colorinlistoftodos]{todonotes}

\newcommand{\const}[1]{\ensuremath{\mathsf{#1}\xspace}}

\begin{document}

\title{Simulating the Effects of Social Presence on Trust, Privacy Concerns \& Usage Intentions in Automated Bots for Finance}

\author{\IEEEauthorblockN{Magdalene Ng\textsuperscript{1*}, 
Kovila P.L. Coopamootoo\textsuperscript{1*}, 
Ehsan Toreini\textsuperscript{1}, 
Mhairi Aitken\textsuperscript{2}, 
Karen Elliot\textsuperscript{2}, 
Aad van Moorsel\textsuperscript{1}}
\textit{\textsuperscript{1}Newcastle University School of Computing, \textsuperscript{2}Newcastle University Business School}, UK \\
\textit{\textsuperscript{*}Corresponding authors: 
magdalene.ng@newcastle.ac.uk \& kovila.coopamootoo@newcastle.ac.uk}
}

\maketitle

\begin{abstract}
FinBots are chatbots built on automated decision technology, aimed to facilitate accessible banking and to support customers in making financial decisions. Chatbots are increasing in prevalence, sometimes even equipped to mimic human social rules, expectations and norms, decreasing the necessity for human-to-human interaction. 
As banks and financial advisory platforms move towards creating bots that enhance the current state of consumer trust and adoption rates, we investigated the effects of chatbot vignettes with and without socio-emotional features on intention to use the chatbot for 
financial support purposes. We conducted a between-subject online experiment with \emph{N} = 410 participants. Participants in the control group were provided with a vignette describing a secure and reliable chatbot called XRO23, whereas participants in the experimental group were presented with a vignette describing a secure and reliable chatbot that is more human-like and named Emma. 
We found that Vignette Emma did not increase participants' trust levels nor lowered their privacy concerns even though it increased perception of social presence.
However, we found that intention to use the presented chatbot for financial support was positively influenced by perceived humanness and trust in the bot.
Participants were also more willing to share financially-sensitive information such as account number, sort code and payments information to XRO23 compared to Emma - revealing a preference for a technical and mechanical FinBot in information sharing.   
Overall, this research contributes to our understanding of the intention to use chatbots with different features as financial technology, in particular that socio-emotional support may not be favoured when designed independently of financial function.  
\end{abstract}


\section{Introduction}
The upsurge in innovative automated designs and internet technology have instigated financial companies to enter a FinTech revolution~\cite{Lai2018ABC, lu2017financial,yang2017uk}, rapidly changing the landscape of our financial industry in particular, disrupting how we perform our financial activities. At the frontend of automated systems, an emergent part of FinTech are virtual assistant technologies. An increasing number of financial businesses are introducing chatbots as one of their services. Their goals and advantages in doing so are many, including \begin{inparaenum}[(1)]
\item to improve customer convenience and enrich quality of service, 
\item to provide a platform for advice, such as in investments, 
\item to lower bank personnel costs, 
\item to help promote their business or other products and 
\item to move towards the direction of alternative and intelligent finance~\cite{Lai2018ABC, lu2017financial}.
\end{inparaenum}
A complementary advancement in FinTech is the concept of \emph{Emotional Banking}~\cite{blomstrom2018emotional}, refering to banks investing and spending resources to investigate their users' feelings about money, subsequently designing processes and products to reflect this understanding. The main aim of emotional banking is to transform the banking culture and develop a better relationship with customers. 

The UK has become a FinTech hub in the last decade~\cite{lu2017financial,yang2017uk}. 
Many successful challenger banks and financial startups are housed in London~\cite{yang2017uk}, such as Cleo, Starling Bank, Monzo, Monese and Transferwise. Challenger banks are distinguishable from traditional brick-and-mortar banks in that they are online-focused 
and place emphasis on FinTech offerings such as chatbots and robo-advisors. The introduction of FinTech has been rapidly changing how people bank and make financial decisions. Many leading traditional banks in the UK and the world have been quick to follow suit, launching their own versions of chatbots as a service for their customers. For example, NatWest and Royal Bank of Scotland launched Cora, as has HSBC and Santander UK. 



While chatbot as a FinTech offering has seen a boom, this is not reflected in research relating to the understanding and examination of financial chatbot use. 
To most people, finance is a sensitive topic~\cite{kaye2014money} and financial information is often not shared lightly. FinBots deal with various personal data that can be sensitive, including financial data. Yet, very little FinTech-related research and chatbot surveys have been conducted here in the UK to our knowledge. As such, the factors that make users trust, disclose to and use FinBots more are largely unknown, 
as are adoption rates of these bots.  Meanwhile, banks and financial advisory platforms all aspire to create banking chatbots or update existing systems to enhance consumer trust and adoption rates. 

We take a leaf from chatbots in other fields that are better-established and researched. A body of literature have successfully applied well-established findings from social psychology to the field of Human-Computer Interaction (HCI) to make chatbots appear more human-like, creating more harmonious interactions with users~\cite{oh2018systematic}. Mimicking human social rules and norms, a series of findings show that chatbots with more human-like features and those that engage in social behaviours gain more positive responses from users~\cite{toader2019effect,looije2010persuasive,dias2013want,araujo2018living,kozima2005using,fong2003survey} 
This design strategy has even been utilised in the area of conflict research, where the virtual agent was designed to appear more human-like and therefore establishing increased rapport with their users~\cite{hasler2014virtual}. 
Concurrently, many of these studies also measure the feeling of being with or interacting with a real human being when using these hyperrealised chatbots. This feeling and perception of interacting with a human being is also known as 'social presence'~\cite{oh2018systematic, widener2019need, ciechanowski2019shades}. 



As banks and financial advisory  platforms are eager to create bots that enhance the current state of consumer adoption rates, we investigate the building blocks of intention to use these bots and information disclosure in the nascent field of financial technology. In particular, in light of botsourcing and chatbots aiming to replace human agents, we postulate that chatbots designed  with social presence features will increase an intention to use the chatbot  for  financial  support purposes.  
In line with previous research findings from chatbots in other areas~\cite{looije2010persuasive, liu2018should}, we propose that socially- and emotionally-apt financial bots will be perceived to be more trustworthy, impacting intention to use the chatbot. We discuss theories pertaining to social presence and social presence features further in Section~\ref{ss:socio-emotional}. 



\begin{RedundantContent}
\paragraph{\emph{Contributions}}
While there is research investigating privacy concerns and trust in human-to-chatbot interaction separately, to our knowledge none have studied the relationship between the two in a financial context with regard to the intention to use chatbots. When included in our regression model, perceptions of social presence, trust and privacy concerns all explain a substantial amount of variance in intention to use FinBots. We draw tentative but important conclusions about how changes in our hypothetical socio-emotional FinBot design, perceptions of trust and privacy concerns are associated with changes in the intention to use FinBots. We further investigate financial actions participants feel comfortable giving control to FinBots to automate, as well as types of information participants are comfortable sharing to FinBots.
\end{RedundantContent}

\paragraph{\emph{Outline}}
We first provide a background section with an overview of the current state of FinBots in the UK. We discuss theories and research involving socio-emotional features in FinBots, together with social presence, trust and privacy concerns in the automated financial sector. 
We then introduce our research aims and describe our methodology.
Next, we present our findings and provide a discussion. Finally, we evaluate our methodology in the limitations section before concluding the paper.

\section{Background}
\subsection{About Chatbots \& FinBots}
A ``bot" is a software application created to automate certain tasks using AI technology. A chatbot is an automated response application that can sometimes be programmed to imitate a real conversation with a human in their natural language ~\cite{Lai2018ABC}. In essence, it is a bot that can chat. Some other names for chatbots are talkbots, chatterbots, IM bots, interactive agents, conversational agents and artificial conversational entities. 

Chatbots have a  machine-learning layer, where the unstructured language input from the user is converted into a structured format on which response decisions can be made. The algorithms can be enabled via auditory or textual interfaces to understand the intent of the user and to send responses. 

One level above this are robo-advisors, which are essentially complex chatbots. Financial robo-advisors are an algorithm-based digital platform that offers automated financial advice or investment management services. As an example, robo-advisors in the asset-management industry are digital investment managers that leverage the Internet to offer customised investment portfolios to clients by employing algorithms.

\emph{FinBot} is a financial chatbot that is programmed to, at the lowest level, digitally assist customers with answers to frequently asked questions and perform simple actions via predefined tree hierarchies such as updating customer's address and checking bank statements. This kind of chabot offers options to customers and will ask them for data. For example, the bot might say ``I am your digital assistant that can help you with all your everyday banking queries, what would you like to do? Select X, Y, Z''. Examples of these button-based chatbots in the UK include Cora (Natwest and Royal Bank of Scotland), Lloyds, Santander, or Aida (Barclays).  

Most if not all UK FinBots are simple chatbots that are programmed to provide customers with guided options to solve simple finance-related queries, all created to increase optimisation and efficiency in banking decisions.
While both traditional and some challenger banks have simple button-based chatbots, many allow customers to connect their bank accounts with more advanced applications like Cleo AI\footnote{meetcleo.com}, Snoop\footnote{snoop.app}and with FinTech brands like TrueLayer. These advanced applications can act as a person's digital financial manager where customers can conduct complex banking procedures such as automated savings, payments, set up budgets and create personalised prompts for bills.

\subsection{Socio-emotional FinBots} 
\label{ss:socio-emotional}
The effort to imbue immersive qualities and human-like behavioural richness in chatbots are in line with the social response theory and the Computers-Are-Social-Actors (CASA) framework. These theories propagate that people use their social norms as a guide when interacting with computers. We perceive computers that greet, make small talk and employ conversational turn-taking to be more reliable, competent and trustworthy. Simply put, humans project expectations of human-to-human interaction onto computers because we have evolved to be socially oriented~\cite{nass1994computers,reeves1996media} 
Several studies show individuals displaying companionable behaviours such as wishing computers `goodnight' and sending it emojis, especially if they perceive that it has a distinct personality~\cite{araujo2018living,nass2000machines,reeves1996media}. In turn, individuals trust and respond well to machines that are socially and emotionally apt in its interaction~\cite{toader2019effect,araujo2018living,looije2010persuasive,dias2013want,kozima2005using,fong2003survey,kidd2008robots,liu2018should,looije2010persuasive}. The perception of human-like behaviours in a computer system has been postulated to combat the initial distrust we have towards machines~\cite{zamora2017m}. 

Because people tend to expect and use human language in their interactions with machines~\cite{Lai2018ABC}, chatbots these days are being programmed to respond with vernaculars, colloquialism, emojis and GIPHYs. The goal is to increase users' comfort, companionship and `click' with the chatbot. If rapport can be achieved through human-like conversational features, this in turn may increase their level of engagement with the chatbot. After all, the ability to converse freely in natural language is one of the hallmarks of human intelligence and considerable efforts have been made in humanising chatbot dialogues across a number of research areas\cite{verhagen2014virtual, nunamaker2011embodied}. These include replicating basic human conversation in the chatbot system, and we discuss some of these behaviours in more depth below. 


\subsubsection{Politeness}
Chatbots can be programmed to take turns in `speaking', a kind of politeness~\cite{reeves1996media}. Another example of politeness is saying `goodbye'~\cite{looije2010persuasive,folstad2018makes}. The association between politeness and trust in chatbots have been qualitatively reviewed~\cite{folstad2018makes} and studied indirectly in other industries such as farming~\cite{jain2018farmchat}. Similarly, a chatbot in the area of health management that observed conversational turn-taking among other social cues was rated as more trustworthy by participants~\cite{looije2010persuasive}. 

\subsubsection{Active Listening}
Active listening facilitates rapport~\cite{von20091} and perceived rapport in turn invites more self-disclosure~\cite{kang2010turn}. If we like a robot that actively responds to us and engages with us~\cite{bigelow2000patient,  kriglstein2005homie}, we trust it more~\cite{kidd2008robots}. 
An example of active listening in human-to-chatbot research is one via self-feeding~\cite{hancock2019learning}. Whenever the bot in this study predicted that the human it was interacting with was unsatisfied via responses such as ``What are you talking about?'', it would ask for feedback by saying things like ``Oops I messed up! What should I have said?''. It then adapted its interactions based on the feedback provided by the user, such as ``Maybe you could have asked me about...''. 

\subsubsection{Empathetic Responses}
A chatbot that displays 
empathethic behaviours and encouraging statements gains more rapport from users, increases self-disclosure~\cite{riek2010my, kang2010turn,devault2014simsensei,kulms2011s}
and it has been incorporated in many companionable bots~\cite{liu2018should,pereira2010using}. As a field example, Cleo AI has been programmed with the ability to motivate and encourage users, using statements such as ``We're going to smash it!'' and ``You had some bank charges recently, but don't worry, I'm coming for your bank'' and ``Here's how well you're doing compared to last month''. 

\subsubsection{Personalisation}
 Personalisation can increase a chatbot's motivational effects~\cite{folstad2019chatbot, prost2013contextualise}. 
 Some examples of personalisation include behaviours such as calling someone by their preferred name (i.e., `Dr White' or, just `James')~\cite{gefen2003managing}, as well as giving feedback~\cite{adam2010hello}. 
Personalisation works as a technique to increase trust because it increases the level of comfort in the interaction, a sense of companionship and therefore coherence and control~\cite{koay2007living}. 
In the FinTech area, Cleo AI has been programmed to personalise her interaction with customers by asking them what she should call them, as well as give personalised feedback based on their financial performance. 

\subsection{Privacy Concerns and Privacy Risks in using FinBots}
Privacy within social interactions is classically viewed as a dynamic process of interpersonal boundary control~\cite{altman1975environment}, involving an interplay between privacy and sharing (as wilful self-disclosure~\cite{jourard1971self}) behaviours. When disclosing information, individuals may struggle to balance conflicting needs of being both open and closed in contact with others~\cite{margulis2003privacy, coopamootoo2017whyprivacy}. 

Privacy concerns and risks when using financial chatbots have not yet been studied. The types of data that needs to be revealed to financial chatbots are personal and can be sensitive, creating concerns of data exposure and rights to data.
When it comes to payment processes involving bank and/or credit card data, data protection is crucial. 

In addition to platform requirements, the General Data Protection Regulation (GDPR) is set as a legal framework for chatbot providers and companies to adhere to within the European Union~\cite{elshekeil2017gdpr}. Areas that require compliance include privacy by design and users' right to erase. While chatbots and robo-advisors need to be GDPR-compliant in the UK, concerns arise despite regulation such as in the field of medical health~\cite{iacobucci2020row}. Privacy concerns have also been reported when using chatbots in other areas and industries, from general smart home assistants to mental health~\cite{kretzschmar2019can, ischen2019privacy}. 
 
Automated advisors raise privacy issues~\cite{vihavainen2013clash} in the form of data breaches, mishandling and misuse of personal data, the inability to regulate what the system monitors and not knowing its intentions for doing so, and being uninformed about the actual workings of the algorithm process ~\cite{michie1994machine}. 
These potential concerns will likely affect user trust and usage of financial chatbots. 
 
The introduction of socio-emotional features in a chatbot may prompt users to share more personal data than necessary for a financial transaction to take place. This can potentially engender a dialectical tension for users. On the one hand, they may fear hackers or organisations gaining unauthorised access to their personal information. On the other hand, they may be instigated with a sharing and connecting attitude with a sociable chatbot. Privacy concerns may take a different shape when using chatbots that comes across as more human-like in its conversations~\cite{ischen2019privacy}. Users may need to perceive chatbots as privacy-friendly to be willing to use it and to be comfortable disclosing information to it. 

\section{The Present Study}
\label{sec:aim}
FinBots (and chatbots in general) deal with various personal and/or sensitive data. Therefore, trust plays a critical role. If trust levels are low, then individuals might not want to use these chatbots for financial purposes nor will they feel comfortable sharing information with it. Insofar, we are not aware of studies investigating the interrelation between perceived social presence, trust, privacy concerns, intention to use, and intention to disclose in the context of financial chatbots. Our aim is to investigate an imagined chatbot equipped with socio-emotional cues and its influence on intention to use and intention to disclose. 
We propose that people will indeed perceive these hypothetical socio-emotional chatbots to be more human-like, and this will positively impact trust and the intention to use this financial bot. We hypothesise that people who perceive the bot as more human-like will also have lowered perceived privacy concerns. 
An increased perception of human-likeness of the chatbot will also increase comfort in intention to share personal and financial information to the chatbot. 

Using a vignette-style between-subjects experimental design, we test five extra features in a simulated chatbot with socio-emotional features not present in the control condition: \begin{inparaenum}[(1)]
\item A human name (i.e., Emma versus XRO23) as an identity cue~\cite{go2019humanizing},
\item the ability to respond empathetically and give encouraging statements~\cite{pereira2010using,kim2018can,bickmore2004towards}, 
\item active listening skills~\cite{kang2012socially,hancock2019learning}, 
\item the ability to personalise, such as using the user's preferred name~\cite{gabbert2020exploring,kanda2007two} and 
\item politeness, including being able to turn-take and make friendly small talk~\cite{nass2000machines,looije2010persuasive,verhagen2014virtual,reeves1996media,kang2010turn}. 
\end{inparaenum}

\subsection{Hypotheses}
\label{sec:hypotheses}
Based on the literature reviewed, we predict that the following hypotheses and investigate their null hypotheses (as provided in Section~\ref{sec:results}): 


\const{H_{1,1}}:
Perceived social presence is significantly different for Vignette Emma compared to Vignette XR023.



\const{H_{2,1}}:
Perceived privacy concern is significantly different for Vignette Emma compared to Vignette XR023.


\const{H_{3,1}}:
Perceived trust is significantly different for Vignette Emma compared to Vignette XR023.

 
\const{H_{4,1}}:
Perceived social presence, trust  and privacy concern have a significant impact on intention to use the imagined chatbots.



\const{H_{5,1}}
Participants feel significantly different in disclosing both financial and personal data to Vignette Emma in comparison to Vignette XRO23. 


\const{H_{6,1}}:
Participants feel significantly different with allowing Vignette Emma to automate financial tasks in comparison to Vignette XRO23.



As a sub-research question, we also want to examine the influence of perceived social presence on privacy concerns and trust perceptions. 
\const{H_{7,1}}: Perceived social presence influence trust and perceived privacy concerns in the imagined chatbots.

\section{Method}
We conducted a between-subjects online study, diligently following the good practice guidelines for empirical research in security and privacy~\cite{coopamootoo2016evidence, coopamootoo2017ifip}, themselves founded on scientific hallmarks. 
\emph{First} we replicated validated methods using the questionnaires described later in Section~\ref{sec:measurement_apparatus}.
\emph{Second}, we defined hypotheses at the fore in Section~\ref{sec:aim} and discussed limitations in Section~\ref{sec:limitations}.
\emph{Third}, we followed the standard APA Guidelines~\cite{american1983publication} in reporting our statistical analyses, effect sizes, assumptions and test constraints. 

\subsection{Participants}
\subsubsection{Recruitment}
We sampled  $\emph{N}=410$ UK participants from Prolific Academic.
The data from Prolific Academic has good quality and good reproducibility compared to other crowd sourcing platforms~\cite{peer2017beyond}.
The study lasted between $20$ to $25$ minutes. Participants were compensated at a rate of \pounds$7.5$ per hour.

\subsubsection{Demographics}
There were $298$ females and $112$ males in our study, totalling to $\emph{N}=410$ participants. The mean age in our sample was $33.12$, $sd=11.71$. After filtering out those who did not complete the study and those who failed attention checks, we were left with $\emph{n}=219$ in the Emma condition and $\emph{n}=191$ in XRO23 condition.

\subsubsection{Assignment}
Participants were randomly assigned to either receive Vignette Emma 
or Vignette XRO23. 
They then answered a battery of questionnaires stated in the subsection below. 

\subsection{Procedure}
This online study consisted of the following materials: \begin{inparaenum}[(a)]
\item a demographic form, 
\item an introduction to either one of the chatbot vignettes,
\item measures of social presence, privacy concerns, trust in chatbot, intention to use chatbot, intention to self-disclose, automation for financial support and
\item two manipulation check items to assess participants' understanding of the vignette.
The experimental design is outlined in Figure 1.
\end{inparaenum}

\begin{figure}[h]
\centering
\includegraphics[scale=0.48]{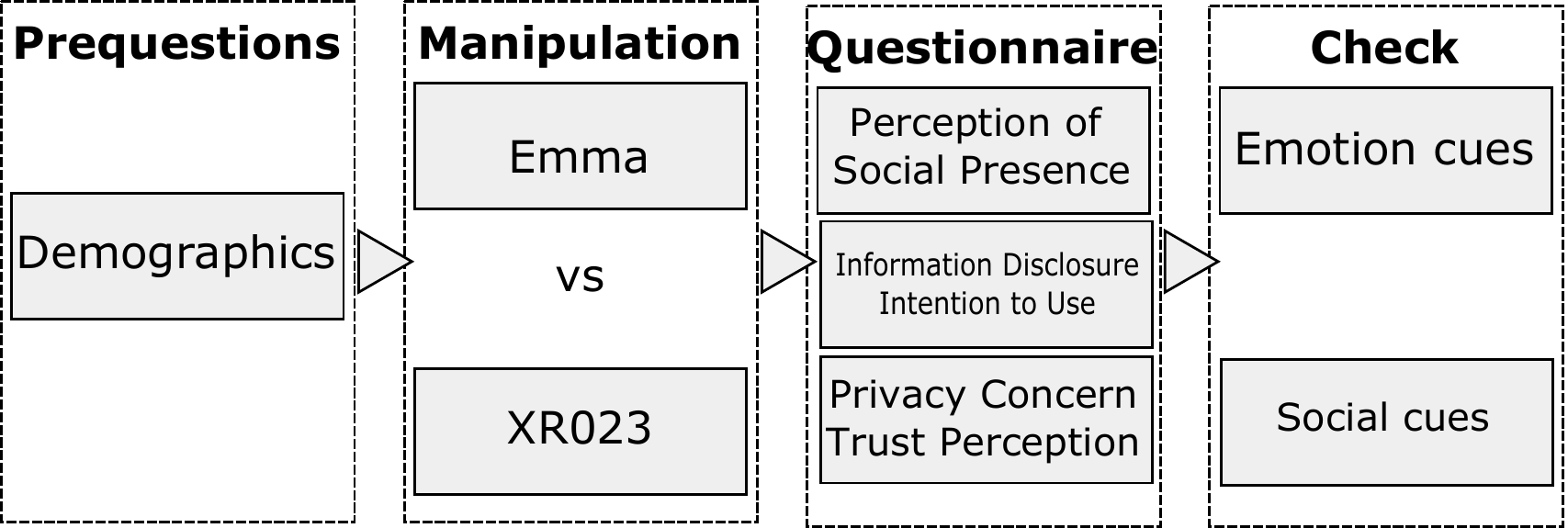}
\caption{Experimental design}
\label{fig:structure}
\end{figure}

\subsection{Stimuli Manipulation}
We opted for a vignette-style methodology as an induction protocol at this stage of our chatbot development. 
Vignette designs are widely used and have advantages~\cite{gould1996using, sleed2002effectiveness}, in particular in enabling the exploration of a situational context and in elucidating influential variables~\cite{barter1999use}. 

The control condition, Vignette XR023, depicted a chatbot that is technical and mechanical. The other condition, Vignette Emma, depicted a socio-emotional chatbot. 
Both chatbots were described as easy to use~\cite{corritore2005measuring}, secure, privacy-focused by design and safe~\cite{folstad2018makes,ischen2019privacy}, financial expert~\cite{verhagen2014virtual,yagoda2012you,folstad2018makes}, speedy~\cite{hancock2019learning}, predictable~\cite{nordheim2018trust, yagoda2012you}, accurate in understanding users' messages~\cite{yagoda2012you}, having a controllable level of automation, and provided by a bank the participants already trust~\cite{nordheim2018trust}. 


\subsubsection{Control condition: Vignette XR023}
Chatbot XRO23 was not given a human name, similar to Araujo's design~\cite{araujo2018living}. 
Vignette XR023 was described as: 
\emph{Imagine a reputable bank that you regularly use have developed a financial text chatbot called XRO23.}

\emph{XRO23 is a financial expert that can assist you. It can analyse your transactions and identifies your regular income, rent, bills and daily spend. Using this and other factors like your available balance, XRO23's algorithm can run every few days and calculates an affordable amount to set aside for you automatically.}

\emph{XRO23 was built with your security and privacy in mind, safely encrypts data and you are in control (i.e., to set more or less money aside).}

\emph{XRO23 is fast in giving you the information you need, saves your time, and is very accurate in understanding the messages you type to it. The relevancy of its content is high. Chatbot XRO23 is easy to use, predictable, flexible and gives quality results.}

\subsubsection{Manipulated condition: Vignette Emma}
We focused on implementing socio-emotional cues that were verbal (textual) and personal in Vignette Emma, disregarding visual and auditory cues as most existing FinBots are text-based. 
Vignette 'Emma' was described as: 
\emph{Imagine a reputable bank that you regularly use have developed a financial text chatbot called Emma.}

\emph{Emma is a financial expert who can assist you. She can analyse your transactions and identifies your regular income, rent, bills and daily spend. Using this and other factors like your available balance, Emma's algorithm can run every few days and can calculate an affordable amount to set aside for you automatically.}

\emph{Emma was built with your security and privacy in mind, safely encrypts data and you are in control (i.e., to set more or less money aside).}

\emph{Emma is polite and can be personalised. She can use your preferred name if you want to, and also understands your feelings when you interact with her. She takes her turn to respond appropriately to the conversation, acknowledges and reacts to your feelings accordingly. She also encourages you when you need it by saying things like ``You're doing great, carry on", or ``Don't be sad, you didn't have better options".}

\emph{Emma is fast in giving you the information you need, saves your time, and is very accurate in understanding the messages you type to her. The relevancy of her content is high. Emma is easy to use, predictable, flexible and gives you quality results.}

\subsubsection{Manipulation Check}
At the end of this study, we checked if participants understood the socio-emotional cues presented in the vignette they were given. The two items were `This chatbot can understand what people are really thinking and feeling' and `This chatbot is aware of what is and is not socially appropriate'. Participants rated both items on a Likert scale ranging from 1 (`Strongly Disagree') to 7 (`Strongly Agree'). 

\subsection{Measurement}
\label{sec:measurement_apparatus}
We introduce and discuss the questionnaires employed in our study below, as well as provide a table of all scales and items in Table~\ref{tab:c_alpha_table} in Section~\ref{sec:results}.
All scales were in 7-point Likert form, ranging from 1 (`Strongly Disagree') to 7 (`Strongly Agree'), unless specified otherwise below.

\subsubsection{Trust}
We used a five-item trust scale by Nordheim~\cite{nordheim2018trust}. The original items were adapted by Nordheim~\cite{nordheim2018trust} from Corritore et al.~\cite{corritore2005measuring} and Jian et al.~\cite{jian2000foundations}. This scale captures the thoughts and feelings of our participants towards the chatbot in their given vignette. It has a reported Cronbach $\alpha$  of $.76$, with a sample item being `I trust this chatbot'.

\subsubsection{Privacy}
We measured privacy concern in FinBots, via a five-item scale by
Ischen et al.~\cite{ischen2019privacy}. We chose this scale because it is already used in a chatbot interaction study, with a reported Cronbach $\alpha$ of $.91$.
Ischen et al.~\cite{ischen2019privacy} adapted this scale from two well-known privacy concern scales, including Xu et al.'s~\cite{xu2008examining} and Van Eeuwen \& Van der Kaap's~\cite{van2017mobile}.  

\subsubsection{Social Perceptions}
\label{sec:social_perceptions}
We measured participants' perceived social presence of the
chatbots in our vignette via a five-item scale from Toader et al.~\cite{toader2019effect}. This contributes to our investigation of how real and human-like participants felt the chatbot to be. This scale reported a Cronbach $\alpha$ of $0.89$, with a sample item being `I felt a sense of human contact when interacting with this chatbot'. 

\subsubsection{Intention to Use}
We measured intention to use these chatbots via a five-item scale by Nordheim~\cite{nordheim2018trust}. The original items were adapted by Nordheim~\cite{nordheim2018trust} from~\cite{zarmpou2012modeling, venkatesh2003user, venkatesh46college}. This scale has a reported Cronbach $\alpha$  of $.96$, with a sample item being `If I have access to chatbots like this I will use it'.

\subsubsection{Intention to Disclose Financial Information}
We asked participants to name what information they would feel comfortable disclosing to the chatbot from the following list: full legal name, date of birth, nationality, current address, account number, sort code, $3$-digit security code at the bank of their bank card, account balance, and their regular payments. Regular payments referred to direct debits, details of standing orders, details of recurring and future related payments, and outgoing or incoming account transactions.

\subsubsection{Automation of Financial Processes}
We further provided participants $12$ financial support actions to select from, with a `Yes' of `No' option. For each of the actions below, participants were asked if they would be comfortable if the chatbot automated this action for them: 
\begin{compactitem}
\item checking their bank balance, 
\item tracking their expenses and spending patterns.
\item making payments and money transfers.
\item reporting their lost or stolen card.
\item acting as their financial manager.
\item making them money-saving recommendations.
\item offering them the ability to automate savings.
\item setting up spending budgets.
\item creating prompts for bills.
\item helping them make financial decisions based on what financial data they give permission for it to analyse.
\item growing their money and make financial investments for them.
\item linking their bank account with the chatbot.
\end{compactitem}


\begin{RedundantContent}
\begin{figure*}
\centering
\includegraphics[scale=0.7]{./figures/experiment_structure_emma}
\caption{Experiment design.}
\label{fig:structure}
\end{figure*}
\end{RedundantContent}

\section{Results}
\label{sec:results}
We provide the scales and their individual items in Table~\ref{tab:c_alpha_table}, as well as the measure of internal consistency via Cronbach $\alpha$, for each scale, as achieved in our study.
We found a high internal consistency within the trust, privacy concern, social presence and intention to use scales, as depicted in Table~\ref{tab:c_alpha_table}.

In this section, we evaluate the manipulation and report on our investigation of the null hypotheses of the predictions made in Section~\ref{sec:hypotheses}.

\begin{table*}[h]
\centering
\caption{Scale Items \& Internal Consistency (Cronbach $\alpha$) achieved in this study.}
\label{tab:c_alpha_table}
\footnotesize
\resizebox{\textwidth}{!}{
\begin{tabular}{llc}
\toprule
\textbf{Scale} & \textbf{Items} & \textbf{Cronbach $\alpha$} \\

\midrule
\multirow{5}{*}{Trust}& I feel that this chatbot is trustworthy & \multirow{5}{*}{.887}\\
& I do not think that this chatbot will act in a way that is disadvantageous to me\\
& I am suspicious of this chatbot\\
& This chatbot appears deceptive\\
& I trust this chatbot\\
&\\

\multirow{5}{*}{Privacy Concerns} & It bothers me when these chatbots ask me for this much personal information &  \multirow{5}{*}{.915} \\
& I am concerned these chatbots are collecting too much personal information \\
& I am concerned that unauthorised people may access my personal information \\
& I am concerned that these chatbots may keep my personal information in an unauthorised way \\
& I am concerned about submitting personal information to chatbot\\
&\\

\multirow{5}{*}{Social Presence} & I would feel a sense of human contact when interacting with this chatbot & \multirow{5}{*}{.930}\\
& Even though I could not see this chatbot in real life, there was a sense of human warmth \\
& If I interacted with this chatbot, I would feel a sense of sociability \\
& If I interacted with this chatbot, I would feel that there would be a person who was a real source of comfort to me\\
& If I interacted with this chatbot, I would feel that there would be a person who is around when I am in need\\
& \\

\multirow{5}{*}{Intention to Use}& If I have access to chatbots like this, I will use it & \multirow{5}{*}{.962}\\
&I think my interest for chatbots like this will increase in the future\\
&I will use chatbots like this as much as possible\\
&I will recommend others to use chatbots \\ 
& I plan to use chatbots like this in the future \\
\bottomrule
\end{tabular}
}
Note: The scales were introduced, together with their sources, in the Measurement Section~\ref{sec:measurement_apparatus}.
\end{table*}

\subsection{Manipulation Check}
We ran Mann-Whitney U tests to determine if there was a significant difference in both our manipulation check items for participants who received Vignette Emma versus those who received Vignette XRO23. 

For the manipulation check item `This chatbot can understand what people are thinking and feeling', there was a significant difference between the two conditions. Participants in the Emma condition scored higher ($Mdn=3$) than those in XR023 condition ($Mdn=2$), with $U=13024.0$, $z=-6.776$, $p<.001$. Similarly, for the manipulation check item, `This chatbot is aware of what is and is not socially appropriate', there was a significant difference between the two conditions. Participants in the Emma condition scored higher ($Mdn=4$) than those in XR023 condition ($Mdn=2$), with $U=12092.5$, $z=-7.498$, $p<.001$.

Our results confirmed success in our manipulation, in that participants in the Emma condition recognised the emotive and social features of the chatbot significantly more than those in the XR023 condition.


\subsection{Differences between Conditions}
We investigated \const{H_{1,0}} that ``There is no significant difference in perceived social presence for Vignette Emma compared to Vignette XR023."
We computed a Mann Whitney U test on perceived social presence between the two conditions.
We found that participants in the Emma condition reported a higher perception of social presence ($Mdn=16.0$) compared to participants in the XR023 condition ($Mdn=10.0$), with $U=13566.0$, $z=-6.162$, $p<.001$.
We reject the null hypothesis \const{H_{1,0}}.
We did not find a statistical difference in privacy concern and perceived trust between conditions, and therefore cannot reject \const{H_{2,0}} and \const{H_{3,0}}.

\subsection{Intention to Use Chatbot}
We computed a linear regression with predictors social presence, privacy concern and trust in bot (and the dependent variable being intention to use chatbot) to investigate \const{H_{4,0}} that ``Perceived social presence, trust and privacy concern do not have a significant impact on intention to use the imagined chatbots." 

Our data met the assumptions of normal distribution and absence of outliers for linear regression computation. The Durbin-Watson statistic for the model is $1.951$, where the required range is between $1.5$ and $2.5$.

\paragraph{\emph{Overall Model}}
The regression model was significant with $F(3,409)=166.716$, $p<.001$ and $R^2=.55$, adjusted $R^2=.549$. The model explains $55\%$ of variability in the outcome variable.

\paragraph{\emph{Effect of Predictors}}
Perceived social presence had a positive effect on intention to use chatbot, $p<.001$, $\beta=.247$. 
Trust also had a positive effect on intention to use chatbot, $p<.001$, $\beta=.792$. 
Privacy concern did not have a significant effect on intention to use chatbot, with $p=.72$. 
We reject the null hypothesis \const{H_{4,0}}. 

\subsection{Financial Information Disclosure and Automation for Financial Support}
\subsubsection{Data Disclosure}
We investigated \const{H_{5,0}} that ``Participants do not feel significantly different in disclosing both financial and personal data to Vignette Emma in comparison to Vignette XRO23". 

We computed a Fisher's Exact test on each of the financial information option in the questionnaire. We found significant differences for disclosure of several financial information items. In particular, significant differences were reported for account number, sort code and payments, with $p < .001$, $p < .001$ and $p = .018$ respectively. There were no significant differences in disclosure of socially-attributed items (such as name, date of birth and address) between the two groups. The full list of items and respective \%s of willingness to disclose data by condition can be found in Figure 2.
We reject the null hypothesis \const{H_{5,0}} for account number, sort code and payments only.

\begin{figure} 
\centering
\includegraphics[height = 10cm, width = .999\columnwidth]{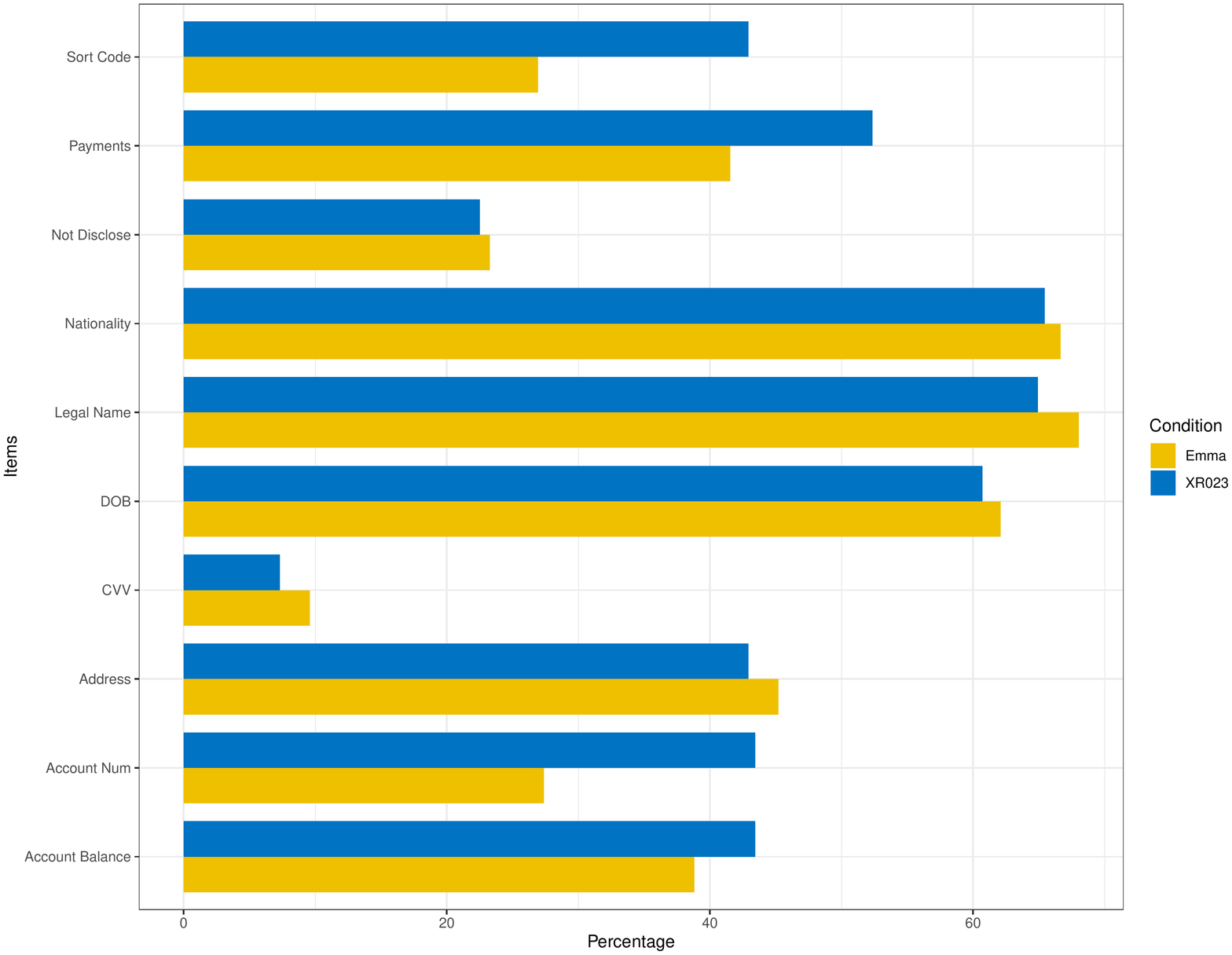}
\caption{Willingness to disclose data items by condition}
\label{fig:data_items}
\end{figure}

\subsubsection{Automation for Financial Support}
To investigate \const{H_{6,0}} that ``Participants do not feel significantly different with allowing Vignette Emma to automate financial tasks in comparison to Vignette XR023", 
we asked participants what financial functions they would feel comfortable for the FinBot to automate for them with a 12-item questionnaire. 
We computed a Fisher's Exact test for each item.
Only three out of these 12 items showed a significant difference between conditions. 
Participants were significantly more comfortable for Emma to make money saving recommendations ($p=.024$) and link their bank account to her ($p=.042$). They were, however, significantly more comfortable for XRO23 to report their lost or stolen card (with $p=.042$). Participants were equally comfortable for both chatbots to check their bank balance, track their expenses and spending patterns, make payments and transfers, set up spending budgets, create prompts for bills, and to make financial investments for them. 
We reject the null hypothesis \const{H_{6,0}} for money saving recommendations, linking bank account to chatbot and reporting lost or stolen card.

\subsection{Further Effects of Social Presence}
As further investigation of social presence and its effects, we investigated \const{H_{7,0}} that ``Perceived social presence does not influence trust and perceived privacy concerns in the imagined chatbots".
We computed two linear regression models with social presence as the predictor, and privacy concern and trust perceptions as dependent variables.

Perception of social presence significantly predicted privacy concern, $F(1, 409) = 45.454$, $p< .001$, where it explained $10\%$ of the variability in privacy concern.
An increase in perception of social presence was associated with a $.283$ unit decrease in privacy concern ($\beta=-.283$).

Perception of social presence significantly predicted trust in chatbot, $F(1, 409) = 91.539$, $p< .001$, where it accounted for $18.3\%$ of the variability in trust in chatbot.
An increase in perception of social presence was associated with a $.348$ unit increase in chatbot trust ($\beta=.348$).
We reject the null hypothesis \const{H_{7,0}}. 

\section{Discussion}
\paragraph{\textbf{Summary of Findings}}
Our study showed no significant difference in trust perception, privacy concern and intention to use between Chatbot Emma and Chatbot XR023.
However, a regression model showed that the predictors social presence and trust influence the outcome variable - intention to use chatbot, explaining $55\%$ of variability, where social presence and trust have a positive and significant effect on intention to use.
We recognise, however, that the `intention to use' items in our questionnaire as provided in Table~\ref{tab:c_alpha_table} did not specify the purpose of use.

In addition, we find that participants were significantly more comfortable with disclosing account number, sort code and payment information to XR023 (FinBot without socio-emotional traits), rather than to Emma. Our vignette design suggests that socio-emotional features in chatbots designed exclusively for automated financial support have little advantage in a FinTech context.

Furthermore, we observed a significantly higher perception of social presence for Vignette Emma compared to Vignette XR023, showing that the different vignettes were successful in inducing the intended difference in condition, although we did not employ a design that allowed actual or continued interaction with the chatbot.
Social presence was also found to significantly influence trust and privacy concern.


\begin{RedundantContent}
\subsection{Perception of Social Presence between Conditions}
Previous studies exploring the influence of interactivity on social presence in virtual reality across a number of fields~\cite{oh2018systematic}, 
showed evidence that people generally feel higher levels of social presence when there is a visual representation available. Prolonged periods of repeated usage and interaction with chatbots~\cite{kidd2008robots, looije2010persuasive} may allow time for initial bias correction~\cite{vanneste2014trust}. 

However, we observed a significantly higher perception of social presence for Vignette Emma compared to Vignette XR023, showing that the different vignettes were successful in inducing the intended difference in condition. --- about needing continued interaction
\end{RedundantContent}

\subsection{Trust and Privacy Concern}
A review of the literature on chatbots in different areas such as the medical field, health management, domestic environments, peace and conflict research show that socio-emotional bots have a positive influence on trust and credibility of chatbots~\cite{liu2018should, kidd2008robots, looije2010persuasive,de2005assessing, hasler2014virtual, fbf4c5a842144422afae9663ed39c5ff}. 
This is reflected in our study, where we do find that social presence has a positive impact on trust and intention to use a FinBot. 

Trust is an important consideration for Artificial Intelligence based systems~\cite{toreini2020relationship}, such as chatbots, and in particular, in the context of finance~\cite{aitken2020establishing}.
Trust has also been reported to be borne out of dynamic interaction over a period of time, enabling error repair~\cite{lewicki2000trust}. 
However, our study design employed vignettes with no interactive component with the financial chatbot. Therefore, it lacks the experiential element where a trusting relationship can be practiced and induced. This would potentially explain the absence of a significant difference in trust responses between conditions. 

We also observed a negative impact of social presence on privacy concern. This can be explained by the depiction of privacy as a metaphorical boundary regulation process, where opening to social connections is related to weaker privacy boundaries and concerns~\cite{derlega1977privacy}.


\subsection{Preference for Chatbot XR023 for Automation of Financial Processes}
Our design isolated socio-emotional and financial features, allowing us to look at the impact of our hypothetical designs on preference for automated financial support. From our results, Chatbot XRO23 may be sufficient for (and even preferred by) users to accomplish their financial goals and tasks. When we looked into previous research on what people look for in a human financial planner in a face-to-face context, Bae \& Sandager ~\cite{bae1997consumers}
found that individuals valued competence, objectivity, the ability to communicate, confidentiality and reliability above other personality traits or personal touches. 
Both vignettes XR023 and Emma in our study have these traits named.

Our results also signal a distinction between personal information and financial information disclosure. While not statistically significant, participants felt more comfortable in disclosing socially-attributed information (such as address, date of birth and full legal name) to Emma. Whereas, they were significantly less willing to give out financial information to her (such as account number, sort code and payment transactions information) compared to XRO23. 

We believe that an interactive design will enable a closer examination of personal information versus financial data sharing to a chatbot, which may involve different aspects of trust and levels of sensitivity. In particular, interpersonal trust for personal information is distinguishable from trust in an institution, merchant and/or chatbot capabilities for financial actions.

\subsection{Emotion in Finance}
Emotional Banking is thought to be one of the biggest trends in FinTech~\cite{blomstrom2018emotional} where retail banks spend time, effort and money on investigating consumers' feelings about their money. Emotional banking involves providing personalised services to customers based on life events and examining how customers respond to products. 

The relationship between emotions and decision-making has been of much interest in psychology for the last several decades~\cite{lerner2015emotion}, now expanding into the area of behavioural finance. 
Some finance related examples also include reports of market investors' behaviours and decisions being influenced by incidental influences such as the weather and their mood~\cite{shahzad2019does}. Studies also show that traders can exhibit irrational and emotionally-led trading decisions in the financial market~\cite{baker2014investor}. 

Albeit, the socio-emotional bot we provided in this study did not make personalised emotional interactions connected to participants' life-events but only introduced a imagined and separate layer of socio-emotional interaction. 
It is likely that if the emotional capability of the bot is integrated with personalised financial services, the findings would be different.

This is further related to the principle of form and function, where Fong et al.~\cite{fong2003survey} asserted that a robot's embodiment should reflect the function of the robot. While we do not use an embodied, human-like financial avatar, the socio-emotional cues embedded in the vignette also did not reflect the financial functions of the chatbot.


\begin{RedundantContent}
Issues: \\
(1) vignette vs software - talking about the socio-emotional capability rather than doing it\\
(2) people may not be conscious of how they feel or how their interaction with bots is impacted by emotion
(3) emotion induction in vignette weaker than in interaction
\end{RedundantContent}

\subsection{Limitations and Future Directions}
\label{sec:limitations}
\emph{Sample.} Although we had a good sample size and used Prolific Academic as online crowd-sourcing platform that provides relative good quality data~\cite{peer2017beyond}, our sample was based on the UK population, yet not strictly representative of the UK population. Inevitably, our sample may represent a portion of individuals who already feel comfortable with interacting and sharing information online. This may have implications for the extent to which their perceptions reflect those in the population. 

\emph{Self-Report.} We relied on participants' responses to an online survey. However, 
we made provisions for the possibility of unreliable samples by introducing several attention checks to establish if our participants had indeed answered our question, before their responses were accepted for further analyses.

\emph{Vignette design.} 
While vignette designs are widely used and are effective~\cite{gould1996using, sleed2002effectiveness}, 
one limitation is that participants can only observe and select cues already presented in the vignette, which, furthermore were only text-based. Trust cues may also manifest visually and even para-verbally. Furthermore, trust in automation changes with user experience of said system~\cite{muir1994trust,bucatariu2017consumer}. 
Because Emma and XRO23 were vignettes, our participants evidently could not test the actual system's accuracy, reliability and predictability. We acknowledge that our variables measured are hypothetical; future directions should consider other mediums besides text-based only (such as image-based) and more interactive methodology (such as avatars) with higher stakes designs. 


\section{Conclusion}
To our knowledge this is a first study investigating
the effects of social presence, privacy concern and trust
in chatbots in a financial context. Our findings depict a tension between
privacy and trust that is brought into focus by the socio-emotional
characteristics of the chatbot, and raises questions
for future research in FinTech. 
We aim to replicate our work with actual chatbot prototypes and with socio-emotional cues integrated with financial decisions. While our focus is on automated bots for finance, we offer a comparison for future work in chatbots in other industries.  

\section{Acknowledgement}
We thank reviewers of EuroUSEC'20 for their feedback. 
This research was funded by the UK EPSRC,  
under grant EP/R033595, ``FinTrust: Trust Engineering for the Financial Industry", \url{https://gow.epsrc.ukri.org/NGBOViewGrant.aspx?GrantRef=EP/R033595/1}.

\balance
\bibliographystyle{IEEEtran}
\bibliography{bibliography}
\balance


\begin{RedundantContent}
\newpage
\section{Response to `Revise' Decisions}
We thank the reviewers for their time and comments, which helped us enhance our paper.

We wanted to make sure we improved the writing of the paper, as this was one of the main comments made by the reviewers. We checked for grammatical errors (i.e., ``which actively responses to us'' changed to ``which actively responds to us'') which previously made sentences unclear in their meaning and/or difficult to read. We carefully proof-read, used more precise language and took care of typos/unusual wordings/missing or hanging sentences. For example, the Introduction section was improved in Paragraphs 2, 3 and 4. 
Other examples - we have improved the writing in the first few paragraphs in the Background section of Chatbots (Section 2.1). We have also rewritten the FinBot section in 2.1 to read better.

We made sure we introduced terminologies and concepts (e.g., challenger banks, social presence, the two bots) clearly this time. A comment from Reviewer B was that ``Social presence is first introduced in the background section, but nowhere in the introduction. It would great to briefly introduce that concept there to show why it comes up later in the background section and study.'' We addressed this by making this terminology clear in the fourth paragraph of the Introduction section, and further expanded this concept in Background section later.

We reviewed relevant theories in Section 2.2, making clear how it relates to the concept of social presence and the other variables in our study (trust and privacy concerns). We were careful with crucial terms such as ``empathetic'' vs. ``empathy'', ``diagonal" vs. ``dialectic" (we took out the words ``diagonal'' and ``empathy'') in Sections 2.2 and 2.3. Reviewer B noted that ``Sub chapter 2.2.2 mixes theoretical background and the implementation in the study''. We have now separated the two and left the implementation of the study to Section 3 in the current version.

We changed the title of Section 2.3 from `Privacy in a Social Context' to `Privacy Concerns and Privacy Risks in using FinBots' to make clear what this section is about. We also improved the writing in this section. 

We introduce the current study in Section 3 in the current version of the paper. We introduced Emma and XR023 before the Hypotheses section (in the second last paragraph), as one of the reviewer's comments was that it was difficult to follow the flow of the paper when these vignettes were introduced only much later. We reformulated the hypothesis as per the comment by Reviewer B (``In terms of RQs it is usually sufficient to state H1'') for an easier read. 

We made clear the differences/similarities between conditions in the last paragraph of Section 4.3. Here, we made clearer descriptions of the two conditions, and made clear the descriptions of similarities/differences between the two conditions.

We removed the conceptual diagram because the meaning was difficult to comprehend in the previous version of this paper. In place, we now made privacy descriptions clearer in Section 2.3. We added the sentence ``Privacy concerns may take a different shape when using chatbots that comes across as more human-like in its conversations. Users may need to perceive chatbots as privacy-friendly to be willing to use it and to be comfortable disclosing information to it''. In Section 3, we also added the sentence ``We hypothesise that people who perceive the bot as more human-like will also have lowered perceived privacy risks and privacy concerns.''

Reviewers A and B found the wording with respect to the statistical analyses for the ANCOVA as well as the wording in the previous version of Section 5.5 (e.g. causality) problematic. We have now removed the ANCOVA and present only regression results in Section 5.3. This makes the results section much clearer to understand and more succinct (to the point), in line with our research questions. We added another regression in Section 5.5 as well, being careful not to denote any causality in both the title/paragraph.

Reviewers pointed out that the section outlining information disclosure to Emma was unclear. We have corrected this, in section 5.4.1. This change is also reflected in our discussion section. In Section 5.4, we added subsections of 5.4.1. (Data Disclosure) and 5.4.2. (Automation for Financial Support) to make it easier to read, when previously these two sections were lumped together.

We lengthened our Discussion section, which was previously only three paragraphs. Reviewer A mentioned that they were unclear about the statement in a previous version of Section 6.1: ``In Emma's paradigm, participants were given the choice to manually override any control they give her. Were subjects in the Emma condition given an option not afforded to those in the XR023 condition?''. In answer to this, yes, both conditions were given this option and we realised including this sentence made it confusing for readers. This sentence has now been taken out. 

We toned down the instances in the paper where it seemed like we had carried out an actual prototype experiment. We had been careful in this version to note in the paper that this is an experiment based on vignettes. We acknowledge and discuss that there no interaction with an actual financial chatbot or a prototype of any sort. We removed any implication of causality (i.e., ``Causal Impact of Social Presence''). 

We discussed our results in comparison to other fields, but it will be difficult to generalise our results to other non-finance contexts some of our measures are finance-specific. Nevertheless, we discussed user perception of chatbots in other areas and industries in both our literature review section and the Discussion section.

Reviewer B commented ``Would it have been an option to provide participants with a fictional chat protocol including real-life examples for financial actions to gain a better insight into how the chatbot would show the features described in the vignette? If that was dismissed as an idea, why?''. In answer to this, a fully interactive chatbot would be the next step of our chatbot development protocol. We rigorously conducted this study with a large sample size as a first investigation. We opted for a vignette here as it is both important and has its advantages for this stage of development. This design allows us to control for chatbot performance factors and enables us to isolate the variables' effects and preclude alternative explanations for our experimental results. We plan to investigate a fully interactive and functioning financial chatbot based on the features inducted in this study and its respective findings.
\end{RedundantContent}
\end{document}